%% file: main.tex
\documentclass[10pt,twocolumn,cleanfoot]{config/asme2e}

\input{config/config}

\input{config/front}

\usepackage{microtype}

\begin{document}
 \setlength{\parskip}{0pt}
 \setlength{\parsep}{0pt}
 \setlength{\headsep}{0pt}
 \setlength{\topsep}{0pt}

% equations
\abovedisplayshortskip=3pt
\belowdisplayshortskip=3pt
\abovedisplayskip=3pt
\belowdisplayskip=3pt

\titlespacing*{\section}{0pt}{18pt plus 1pt minus 1pt}{3pt plus 0.5pt minus 0.5pt}

\titlespacing*{\subsection}{0pt}{9pt plus 1pt minus 0.5pt}{1pt plus 0.5pt minus 0.5pt}

\titlespacing*{\subsubsection}{0pt}{9pt plus 1pt minus 0.5pt}{1pt plus 0.5pt minus 0.5pt}

\microtypesetup{nopatch=item}
\maketitle
\microtypesetup{patch=item}

%---------------
\begin{abstract}\noindent
\textit{%
Route planning for military vehicles is a complex decision-making problem due to the simultaneous influence of environmental trafficability and tactical risks. This paper presents an optimization model that integrates soil trafficability and risk of enemy engagement into a decision-support model for planning activities in open terrain.  Although a military application is the focus of this paper, other use cases include wildfire response, agricultural operations, and off-road vehicle recreation.
The routing problem is formulated as a minimum cost mixed-integer linear program over a discretized representation of the operational environment. Each node represents a location and is connected by arcs to adjacent nodes whose traversal incurs a cost derived from a composite risk function that accounts for soil strength and the proximity to known enemy activity and prior convoy routes. Environmental inputs required for evaluating soil strength are obtained by integrating external models, which estimate spatial variations in the rating cone index (RCI) across the terrain. 
The model is evaluated through a case study conducted at a location in northern Colorado using fine-resolution environmental data and simulated tactical conditions. Scenario analyses demonstrate how variations in risk weighting, vehicle mobility characteristics, and operational conditions influence route geometry and mission risk.  The objective function values achieved varied by five orders of magnitude based on the coefficients assigned to the terms in the cost function and the vehicle properties of the scenario. The results illustrate the capability of the proposed framework to quantify trade-offs between environmental mobility constraints and tactical considerations.%
}
\end{abstract}

\vspace{1ex}

% suggestions (not in a specific order):
% control co-design
% floating offshore wind turbines
% linear parameter-varying models
% optimization?
% optimal control?
% renewable energy systems?
% active system design?

%---------------
% \clearpage
\input{input/introduction.tex}
%---------------
% \clearpage
\input{input/Section2}
%----------------
% \clearpage
\input{input/Section3}
%---------------
% \clearpage
\input{input/Section4}
%---------------
%\clearpage
\input{input/Section5}
%---------------
%\clearpage
\input{input/conclusion.tex}
%--------------
%\clearpage
\begin{acknowledgment}
This research was partially sponsored by the Army Research Laboratory and was accomplished under Cooperative Agreement Number W911NF-24-2-0179.  The views and conclusions contained in this document are those of the authors and should not be interpreted as representing the official policies, either expressed or implied, of the Army Research Office or the U.S. Government.  The U.S. Government is authorized to reproduce and distribute reprints for Government purposes notwithstanding any copyright notation herein.
The authors would also like to thank Dr. Joseph Scalia for help in interpreting the results of the STRESS model and tailoring it to provide input for the optimization model. 
\end{acknowledgment}
%--------------------------
% \clearpage
\renewcommand{\refname}{REFERENCES}
\bibliographystyle{config/asmems4}
\begin{mySmall}
%\nocite{*} % remove later, displays all references
\bibliography{References}
\end{mySmall}

%--------------------------
% \clearpage
% \onecolumn
% \xneed{Will be removed in the final version.}
% % This will be removed in the final version.
% \tableofcontents
% \listoffigures
% \listoftables
% -----------

%---------------
%  \clearpage
% \input{input/working.tex}

\end{document}

%% file: config/config.tex
\usepackage{amssymb}

\usepackage{lmodern}
\usepackage{amsmath}
\usepackage{amsfonts}

\usepackage{bm} % bold for greek symbols

\usepackage{txfonts}

\usepackage[T1]{fontenc}
\usepackage[utf8]{inputenc} % accept different input encodings

\usepackage[dvipsnames]{xcolor} % defines colors

\usepackage[american]{babel} %  culturally-determined typographical rules

\usepackage{graphicx}

\usepackage[noadjust]{cite} % grouped citations

\usepackage{mathtools}

\usepackage{subcaption}% for subfigures

\usepackage{hyperref}

\usepackage{enumitem}

\usepackage{flushend} % add at the end

\usepackage{titlesec}

\input{config/doiCmd.tex}

\newcommand{\xsection}[1]{\section[#1]{\MakeUppercase{#1}}}

%% file: config/doiCmd.tex
\usepackage{hyperref}

\newcommand{\doi}[1]{{doi:~\href{https://doi.org/#1}{#1}}\rmFullStop}

\newcommand*{\rmFullStop}{\rmifnextchar{.}{}{}}

\makeatletter
\newcommand{\rmifnextchar}[3]{%
  \begingroup
  \ltx@LocToksA{\endgroup#2}%
  \ltx@LocToksB{\endgroup#3}%
  \ltx@ifnextchar{#1}{%
    \def\next{\the\ltx@LocToksA}%
    \afterassignment\next
    \let\scratch= %
  }{%
    \the\ltx@LocToksB
  }%
}
\makeatother

%% file: config/front.tex
\confshortname{\phantom{IDETC/CIE2026}}

\conffullname{\phantom{Proceedings of the ASME 2026} \\  \phantom{International Design Engineering Technical Conferences and}\\ \phantom{Computers and Information in Engineering Conference}}

\confdate{\phantom{23-26}}
\confmonth{\phantom{August}}
\confyear{\phantom{2026}}
\confcity{\phantom{Houston, TX}}
\confcountry{\phantom{USA}}
\papernum{\phantom{DETC2026-188697}}

\title{Multi-Criteria Integer Programming Model for Route Planning in an Off-Road Combat Environment}

\author{Joshua Betz
\affiliation{%
Graduate Student\\
Systems Engineering \\
Colorado State University \\
Fort Collins, CO 80523 \\
\texttt{\href{mailto:joshua.betz@colostate.edu}{joshua.betz@colostate.edu}}
}
}
\author{Daniel~R.~Herber
\affiliation{%
Associate Professor\\
Systems Engineering \\
Colorado State University \\
Fort Collins, CO 80523 \\
\texttt{\href{mailto:daniel.herber@colostate.edu}{daniel.herber@colostate.edu}}
}
}
\author{Jeffrey D. Niemann
\affiliation{%
Faoro Professor\\
Civil \& Environmental Engineering \\
Colorado State University \\
Fort Collins, CO 80523 \\
\texttt{\href{mailto:jeffrey.niemann@colostate.edu}{jeffrey.niemann@colostate.edu}}
}
}

\hypersetup{
    unicode=false,          % non-Latin characters in Acrobat’s bookmarks
    pdftitle={Multi-Criteria Integer Programming Model for Route Planning in an Off-Road Combat Environment},    % title
    pdfauthor={Joshua Betz and Daniel R. Herber and Jeffrey D. Niemann},     % author
    pdfsubject={},   % subject of the document
    pdfkeywords = {}, % list of keywords
    pdfnewwindow=true,      % links in new window
    colorlinks=true,
    allcolors=blue
}

%% file: input/introduction.tex
\xsection{Introduction}
\label{sec:introduction}
The modern battlefield is a dynamic and multi-domain environment.  Maneuvering ground forces through such an environment involves significant uncertainty and risks from many sources, which makes it difficult to traverse even in the most favorable conditions.  Furthermore, in the era of highly advanced and mechanized military forces, effective movement planning throughout the battlefield can be a decisive factor in determining mission success or failure. Due to the complex nature of modern warfare, many (often competing) considerations must be taken into account when planning to maneuver within a combat environment, and infinite potential routes exist in a continuous plane.  Thus, developing a methodology to quantify and make recommendations for risk-based decisions concerning route planning on the battlefield would help achieve more consistent and improved mission outcomes. Although previous research presents models that can assist with route planning in open terrain, none of them include tactical considerations that a battlefield commander must include in their planning \cite{talhofer2016, pundir2022, tarapata2003, mccullough2017, priddy1995, pokonieczny2023}.

%\subsection{Terrain Considerations}
Terrain trafficability depends on environmental characteristics (such as soil strength and topographic slope) and vehicle characteristics (such as weight and tread properties).  Because military vehicles are often equipped with armor and other specialized equipment, they are typically much heavier than civilian vehicles; therefore, much greater soil strength is required for traversing off-road. Soil strength depends on the composition or texture of the soil as well as the soil moisture.  Due to the spatial heterogeneity of soils and the space-time variability of weather patterns, substantial uncertainty is inherent in estimates of soil strength. 

%\subsection{Tactical Considerations}
When assessing vehicle routing in open terrain, it is essential to consider the tactical implications of a route to minimize the kinetic risk to troops. Toward this end, friendly, enemy, and civilian activity is important to analyze. Recent conflicts have shown that enemy combatants are likely to stage weapons and personnel in areas with civilian populations, so avoiding civilian populations can minimize risk to troops as well as civilians and friendly forces. Intelligence gathering may also indicate areas of increased enemy activity, which should be considered when route planning.  Finally, past vehicle routes will be known to enemy combatants, so avoiding past travel paths will lower the enemy's ability to stage attacks and employ explosive hazards.

Open-terrain route optimization also has broad relevance across multiple civilian domains where movement occurs outside structured transportation networks. For example, in wildfire response, optimizing the routes of ground crews and heavy equipment can reduce response times, improve firefighter safety, and minimize exposure to hazardous conditions such as steep slopes, weak soils, and rapidly changing fire behavior. In agriculture, route optimization can support the movement of heavy machinery in and between areas of production by reducing the risk of operational delays from equipment immobilization and yield losses from crop and soil damage. In addition, off-road recreational travel (such as all-terrain vehicle use) can benefit from risk-aware routing that balances environmental protection and user safety. In each of these contexts, integrating environmental data with operational constraints to formulate quantifiable risks enables more informed decision-making.

The primary objective of this project is to formulate an optimization model that utilizes soil strength data derived from external sources to optimize potential routes through open terrain, where the objective function minimizes risk to mission success associated with getting stuck, maneuvering over unfavorable terrain, and tactical considerations. 

%% file: input/Section2.tex
\xsection{Literature Review}
\label{sec:Lit Review}

Several models have been developed to assess vehicle mobility.  One of the most prominent is the NATO Reference Mobility Model (NRMM).  This model was initially developed in the 1960's and 1970's by the US Army Tank Automotive, Research, Development, and Engineering Center \cite{bradbury2016}.  Since then, the model has been progressively modernized \cite{priddy1995, bradbury2016, mccullough2017}.  NRMM uses empirical relationships to model the interactions between the soil and vehicle and to determine the likelihood that a vehicle can safely pass through terrain and the vehicle's expected speed.  NRMM had many limitations in its early iterations.  One significant limitation was a lack of documentation for the test data and empirical relationships that were used because they were developed over many years by several researchers who eventually left the development program \cite{bradbury2016}.  Also, the model is heavily dependent on empirical observations, so predictions of vehicle behavior beyond the test conditions may not be accurate \cite{bradbury2016, mccullough2017}.  Because of these limitations, NRMM has been used to design trade studies for new vehicles or modifications where the evaluation only considers specific scenarios within the operational limits of the model \cite{bradbury2016}.  Although NRMM has been widely used in military settings, the output of the model only describes where a vehicle can go with minimal risk of getting stuck, not where it should go based on the complete operational picture.  This limitation is important because there are still many tactical considerations that must be made when route planning.  

Shortfalls in NRMM's representation of ground-vehicle interactions led to the recent development of the Next Generation - NATO Reference Mobility Model (NG-NRMM), which includes the following upgrades:
\begin{enumerate}
    \item Geographical Information System terrain data as an input to mobility simulation software
    \item A physics-based terramechanics model that simulates the interface between the vehicle and terrain
    \item Tools to provide terrain parameters needed by the terramechanics model
    \item Stochastic treatment of terrain and vehicle data \cite{mccullough2017}
\end{enumerate}
These improvements allow for a wider range of applications, but the output still aims to identify where a vehicle can go, not where it should go.  

Optimization has been used for vehicle routing in open terrain \cite{tarapata2003, pokonieczny2023, khatiwada2023, talhofer2016, pundir2022}.  Oftentimes, network optimization structure is used due to its computational efficiency and the ease with which terrain can be discretized to assign costs for traveling between nodes of a network.  Once the graph is constructed, there are several methods for solving the optimization problem.  Two common methodologies are graph algorithms, such as Dijkstra's Algorithm or integer optimization (by formulating the problem as a shortest path network) \cite{tarapata2003, pokonieczny2023}.  Dijkstra's algorithm is used to find the shortest path through a network, but its complexity is $O(n^2)$ where \textit{n} represents the number of nodes present in the graph.  Thus, decreasing the size of the nodes or increasing the planning area has a major impact on the computation time of the model \cite{griva2008}.  However, there are many variants of Dijkstra's Algorithm that may be more suitable for the problem at hand to increase computational efficiency \cite{tarapata2003}.  However, the studies presented by Tarpata and Pokonieczny both acknowledged limitations in their methodology because suitable terrain is only one factor that should be considered in the optimization.  In order to overcome the limitation of only considering the terrain, Tarpata suggested that their model be used in conjunction with a tactical analysis tool, while Pokonieczny identified multiple pathways through the terrain while classifying them based on terrain suitability.  The commander could then select a route from the collection of pathways based on other tactical considerations \cite{tarapata2003, pokonieczny2023}.  However, neither of these studies quantified the contribution of tactical considerations relative to terrain considerations to identify an optimal pathway.

%% file: input/Section3.tex
\begin{figure}[t] %p = figure gets its own page
\centering
\includegraphics[scale = 1, trim = {0.5in 0 0.5in 0}, clip]{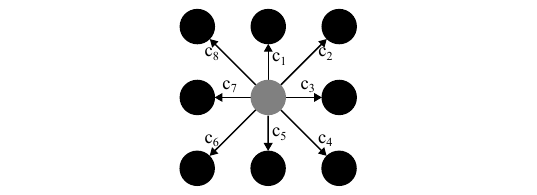}
\caption{Cost arcs between adjacent nodes}
\label{fig:nodes}
\end{figure}

\xsection{Methodology} \label{sec:methodology}
In this study, a route is obtained through the solution of a mixed-integer linear program that is formulated as a minimum cost flow network where the mission area is discretized in a grid pattern and traversing nodes incurs a cost that is derived from the risk associated with occupying the area inside the node.  The risks considered in the formulation include environmental risks associated with traversing soil that has insufficient estimated strength to bear the weight of the vehicles in the convoy, as well as tactical risks originating from traveling where other convoys have been recently and traveling near areas of known enemy activity.  These risks are weighted in the formulation to normalize their contribution to the overall cost of traversing an arc connecting the nodes, making the cost unitless.  Figure~\ref{fig:nodes} shows that from each node, there is an associated arc in every adjacent direction with a unique cost associated with traversing it.

\begin{figure}[t] %p = figure gets its own page
\centering
\includegraphics[scale = 0.99]{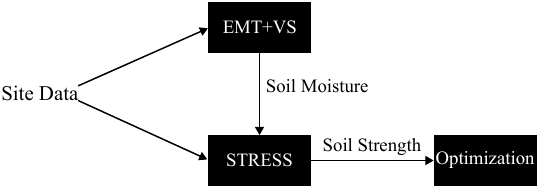}
\caption{Modeling architecture}
\label{fig:models_figure}
\end{figure}

%\subsection{Site Data Procurement}
Two additional external models are used to estimate the soil strength, which can vary in space and time.  First, the Equilibrium Moisture from Topography Plus Vegetation and Soil (EMT+VS) model is employed to estimate the soil moisture across the area of operations \cite{fischer2025}.  Next, the estimated soil moisture is used in conjunction with soil composition information to estimate the soil strength using the Strength of Surface Soils (STRESS) model.  The STRESS model produces estimates for the friction angle and the moisture variable cohesion across the area of operations.  These soil strength properties are then combined to produce the rating cone index (RCI), which is a single empirical measure of soil strength  \cite{Pauly2019}. 
% The RCI is used in the optimization.
Figure~\ref{fig:models_figure} shows the inputs and outputs of each model used in this study.  It should be noted that the optimization algorithm can readily be coupled with other models of soil moisture and strength.  

%\subsection{EMT+VS Model}
Soil moisture is a critical factor in determining soil strength and vehicle trafficability. Although gridded, near real-time soil moisture data can be obtained from remote sensing (e.g., NASA's Soil Moisture Active Passive satellite), but the spatial resolution of these estimates is far too coarse for mobility assessments (e.g., 9-km grid cells).  Downscaling is required to reach more usable fine resolutions (e.g., 10 m grid cells) .  Downscaling techniques use fine-resolution data for related variables to infer the fine-resolution variations of soil moisture.  Numerous downscaling methods are available, but the EMT+VS model is selected because it reaches finer resolutions than many downscaling methods, and it has been tested for a variety of regions.  The EMT+VS model downscales soil moisture using a water balance for the surficial soil layer (top 5 to 30 cm of the soil) \cite{coleman2013,ranney2015}.  Four processes are considered in the water balance, including infiltration of water into the soil, drainage to deeper soil layers, lateral flow, and evapotranspiration to the atmosphere \cite{hoehn2017,coleman2013,ranney2015}.  The magnitude of each process in each grid cell is inferred from fine-resolution topographic characteristics, which are derived from a digital elevation model, and fine-resolution vegetation characteristics, which are derived from remotely sensed multispectral data.  Unlike conventional hydrologic models, the EMT+VS model does not iterate through time.  It accepts coarse-resolution soil moisture on a given date and produces fine-resolution soil moisture for that same date, without considering past conditions.

Because the EMT+VS model was initially formulated and tested for small regions, Cowley et al. generalized the model to consider regions with large elevation ranges, which might exhibit orographic variations in precipitation and evapotranspiration \cite{cowley2017}.  The model was subsequently adapted to consider multiple coarse resolution grid cells, which allows its application to larger regions \cite{hoehn2017}.  Further development took place to add parameter estimation methods \cite{grieco2018}, and the capability to simulate stochastic variations in soil moisture if desired \cite{deshon2020}.  Most recently, Fischer et al. evaluated the EMT+VS model's accuracy when it uses various sources of input data and showed that the EMT+VS model is more accurate than two other downscaling methods that rely on similar inputs.  This study uses the EMT+VS model version presented by Fischer et al. \cite{fischer2025}.

%\subsection{STRESS Model}
The STRESS model applies a physics-based approach to model the unsaturated soil mechanics \cite{Pauly2019, bullock2024}.  For inputs, it requires soil moisture (which can be obtained from the EMT+VS model or other sources) and soil texture in the form of percent sand and percent clay (which can be obtained from soil maps). The percent sand and clay are used in the ROSETTA model \cite{schaap2001} to determine the soil hydraulic characteristics (i.e., the soil's moisture retention properties). The friction angle and apparent cohesion of the soil (i.e., basic soil mechanical properties) are then determined based on the soil texture classification.  Next, the unsaturated soil mechanics model developed by Lu et al. \cite{lu2013} is used to determine the moisture variable cohesion, which accounts for the soil moisture.  The friction angle and moisture variable cohesion are key physics-based measures of soil strength and can be used in the NG-NRMM model.  However, more empirical terramechanics models (such as NRMM) require the empirical RCI.  In the STRESS model, the RCI is estimated from the physics-based strength characteristics using a method from Perkins et al. \cite{perkins1992}.
The RCI can then be compared to the vehicle cone index (VCI) of a vehicle of interest to determine whether the soil can support that vehicle.

\subsection{Optimization Model Description}

In the optimization model, let $\mathcal{N}$ represent the nodes for the discretized locations as determined by the EMT+VS and STRESS models, where $n\in \mathcal{N}$ represents a generic node in this set.
Similarly, there are additional nodes for the source and sink, which are represented by $s \in \mathcal{S}$ and $t \in \mathcal{T}$, respectively.  The only arc that exists from the source is a cost-free arc that connects the source and the user-defined starting node.  Similarly, the only arc that exists to connect the sink node is a cost-free arc that connects the sink and the user-defined end node.  All other nodes are connected to each of their neighboring nodes by arcs that have a cost of traversal, $c_{i,j}$, derived from the cost function in Equation~\eqref{eq: cost} based on the risks associated with getting stuck in the terrain and enemy activity.  The decision variable $X_{i,j}$ is a binary variable that takes the value of 1 if the optimal route includes traversal between nodes $i$ and $j$, and 0 otherwise.

              %%%%%%%%%%%%%%%%%%%%%%%%%%%%%%%%%%%%%%%%%%%%%%%%%%%%%%%%%%%%%%
              %%%%%%%%%%%%%%%        MODEL FORMULATION       %%%%%%%%%%%%%%%
              %%%%%%%%%%%%%%%%%%%%%%%%%%%%%%%%%%%%%%%%%%%%%%%%%%%%%%%%%%%%%%
%\subsection{Model Formulation}
                     %%%%%%%%%%%%%        SETS       %%%%%%%%%%%%%
% \[
% \begin{array}{llll}
% n          &\in&    \mathcal{N}              &       \mbox{Set of all nodes on the map}\\
% s          &\in&    \mathcal{S}              &       \mbox{Set containing the source node}\\
% t          &\in&    \mathcal{T}              &       \mbox{Set containing the sink node} \\ 
% i,j      &\in&    \mathcal{A}              &       \mbox{Set of all arcs connecting nodes $i,j$} 
%  \hspace*{10in}
% \end{array}
% \]
%                    %%%%%%%%%%%%%        PARAMETERS       %%%%%%%%%%%%%
% \[
% \begin{array}{llll}
% c_{i,j}    &  \mbox{Cost of traversing arc $i,j$}  &&       \left[ -  \right]     
% \hspace*{2in}
% \end{array}
% \]  
%
                    %%%%%%%%%%%%%        VARIABLES       %%%%%%%%%%%%%

                %%%%%%%%%%%%%        OBJECTIVE FUNCTION       %%%%%%%%%%%%%

The optimization problem is now written as a minimum cost network linear program:
\begin{subequations}
\begin{align}
\text{changing:} \quad & X_{i,j} \\
\text{minimize:} \quad & \sum_{(i,j) \in \mathcal{A}} c_{i,j}X_{i,j}  \\
\text{subject to:} \quad & \text{Equations~\eqref{Continuity}--\eqref{Sink}}
\end{align}
\end{subequations}

\noindent with $(i,j) \in \mathcal{A}$ indicating that an arc exists between node $i$ and node $j$ in the set of all arcs.

                 %%%%%%%%%%%%%        CONSTRAINTS       %%%%%%%%%%%%%
Now, enumerating the constraints first, Equation~\eqref{Continuity} is a flow balance constraint that requires the flow into every node to equal the flows out of the node:
\begin{equation} \label{Continuity}
    \sum_{(i,k) \in A} X_{i,k} = \sum_{(k,j) \in A} X_{k,j} \quad \forall k \in \mathcal{N} \neq S, T
\end{equation}

Next, Equation~\eqref{Source} requires that the arc from the source node to the starting node specified by the user is used in the solution:
\begin{equation} \label{Source}
    \sum_{(s,j) \in \mathcal{A}} X_{s,j} = 1
\end{equation}

\noindent This constraint is vital to the formulation because the source node supplies the network with the initial flow that will propagate through the network in order to satisfy the flow balance constraint in Equation~\eqref{Continuity}.

Similarly, Equation~\eqref{Sink} ensures that the flow enters the sink node at the terminal location of the convoy specified by the user:
\begin{equation} \label{Sink}
    \sum_{(i,t) \in \mathcal{A}} X_{i,t} = 1
\end{equation}

              %%%%%%%%%%%%%%%%%%%%%%%%%%%%%%%%%%%%%%%%%%%%%%%%%%%%%%%%%%%%%%
              %%%%%%%%%%%%%        PARAMETER DERIVATIONS       %%%%%%%%%%%%%
              %%%%%%%%%%%%%%%%%%%%%%%%%%%%%%%%%%%%%%%%%%%%%%%%%%%%%%%%%%%%%%
%\subsection{Cost Parameter Derivation}
The sources of risk that are being considered inside this model are all proportional to the relative risk of occupying a node and thus can be combined into one term using all applicable factors.
Below is the derivation of the cost parameter $c$ (noting that the subscripts $(i,j)$ are omitted for clarity) with each of its constituent terms that is applied to each node:
\begin{equation} \label{eq: cost}
    c = p + h + r + \mu
\end{equation}

\noindent where the applicable risk factors are:
\begin{enumerate}[nosep]
    % \item \textbf{Total Cost, c:}\\
    % This is the cost associated with occupying a node based on the risk of all applicable factors.
    \item \textbf{Soil strength, $p$:}
    \begin{equation}
            p = 
      \begin{cases}
           P ( VCI_{50}-RCI), \; \text{if $VCI_{50} \geq RCI$}\\
           0, \; \text{otherwise}
       \end{cases} 
    \end{equation}
    
\noindent where:
\begin{enumerate}[nosep,label={\textit{(\alph*)}}]

\item $P = $ Risk coefficient associated with the soil strength

\item $VCI_{50} =  $ Vehicle Cone Index of the limiting vehicle type in the convoy for 50 passes

\item $RCI   =  $ Rating Cone Index of the soil inside the node 

\end{enumerate}

This type of comparison is what the Next Generation Mobility Model is based upon \cite{FM5}.

\item \textbf{Recent convoys, $h$:}
\begin{equation}
h = H\exp{-\frac{d_h}{k_h}} 
\end{equation}
\noindent where:
\begin{enumerate}[nosep,label={\textit{(\alph*)}}]

\item $H = $ Risk coefficient associated with occupying the name node as a previous convoy

\item $d_h = $ Distance between the node being considered and the closest node used by any previous convoy

\item $k_h = $ Exponential decay constant

\end{enumerate}

\item \textbf{Known enemy activity, $r$:}
\begin{align}
r = R\exp{-\frac{d_r}{k_r}} 
\end{align}
\noindent where:
\begin{enumerate}[nosep,label={\textit{(\alph*)}}]
\item $R = $ Risk coefficient associated with occupying the name node as previous enemy activity 
\item $d_r = $ Distance between the node being considered and the closest instance of previous enemy activity
\item $k_r = $ Exponential decay constant

\end{enumerate}

% \clearpage
\item \textbf{Small number, $\mu$:}
\begin{equation}
    \mu = 0.1
\end{equation}
\noindent where this penalty is present to ensure that every node has a positive risk value.  This term is important because for nodes with sufficient soil strength that are far away from enemy activity and historical routes can cause unnecessary terrain traversal due to a lack of penalty associated with it.  This number is small enough not to dominate 
% any of the 
other penalties, but large enough to ensure that there is no unnecessary routing.

\end{enumerate}

%% file: input/Section4.tex
\begin{figure*}[t] %p = figure gets its own page
\centering
\includegraphics[scale = 0.7]{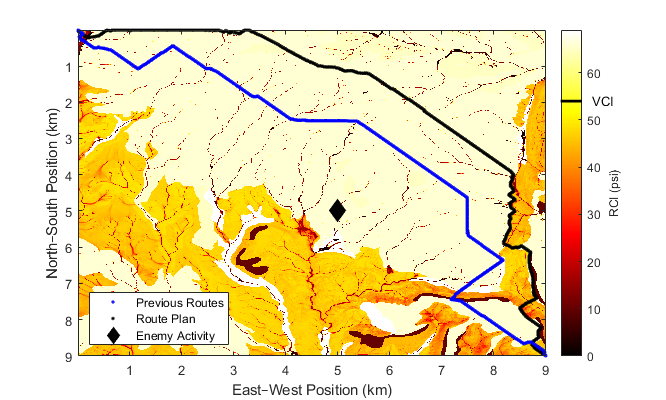}
\caption{Solution given from model execution using nominal input parameters}
\label{fig:Results_Run1}
\end{figure*}

%------------------------------------------
\xsection{Results}\label{sec:Results}
The region used for development and testing of the route planning model is Maxwell Ranch, which has previously been used for testing the EMT+VS and STRESS models \cite{fischer2025, bullock2024}.  Maxwell Ranch is an active cattle ranch that spans approximately 4,000 ha and is situated 53 km northwest of Fort Collins, CO.  The ranch is in the foothills of the Front Range.  It is primarily grasslands with areas of shrubs and widely-spaced evergreen trees.  The bedrock is mainly granite, and the soil is primarily gravelly sandy loam, but peat occurs near streams in some valley bottoms \cite{bullock2024, fischer2025}.  To apply the EMT+VS model, the topography was characterized using a 10-m digital elevation model, and the vegetation was characterized using 10-m multispectral data from the Sentinel-2 satellite.  Only one date, June 10, 2022, is considered in this study, which is a wet date due to a recent precipitation event.  These datasets were supplied to the EMT+VS model, which was used to simulate the top 5 cm of the soil.  The model's parameters were calibrated by comparing the model predictions to in situ measurements of soil moisture at 86 locations on 10 dates.  Ultimately, the EMT+VS model estimates soil moisture at 10 m resolution, and this dataset is then supplied to the STRESS model.  The stress model combines the soil moisture with soil texture information from the Soil Survey Geographic (SSURGO) database (Soil Survey Staff, 2023) to determine the RCI. 
The map of RCI values from the STRESS model is shown in Figure~\ref{fig:Results_Run1}. In order to produce a more challenging mobility environment (which produces more insightful testing of the optimization model), the RCI values were reduced by a factor in Figure~\ref{fig:Results_Run1} (and subsequent analyses). The lowest RCI values occur in the valley bottoms.  These locations tend to be wetter due to the movement of water from the surrounding hills towards the valley bottoms.  The increased soil moisture in the valleys produces weaker soils.

% \begin{figure*}[t] %p = figure gets its own page
% \centering
% \includegraphics[scale = 1]{Figures/Results_STRESS.png}
% \caption{Stress model results}
% \label{fig:STRESS_Results}
% \end{figure*}

\begin{table}[b]
\centering
\renewcommand{\arraystretch}{1.2}
\caption{Key parameter values used in the optimization model}
\label{tab:Parameters_Table}
\begin{tabular}{ccc}
\hline\hline
\textbf{Parameter}            & \textbf{Value} 
& \textbf{Units}
\\ \hline
\textbf{$VCI_{50}$}   & 54   & [psi]            \\ \hline
\textbf{$P$}          & 5    & [-]          \\ \hline
\textbf{$H$}          & 20   & [-]          \\ \hline
\textbf{$k_h$}        & 15   & [-]          \\ \hline
\textbf{$R$}          & 20   & [-]          \\ \hline
\textbf{$k_r$}        & 30   & [-]          \\ \hline\hline
\end{tabular}
\end{table}

\begin{figure*}[p] %p = figure gets its own page
\centering
\includegraphics[width = \textwidth]{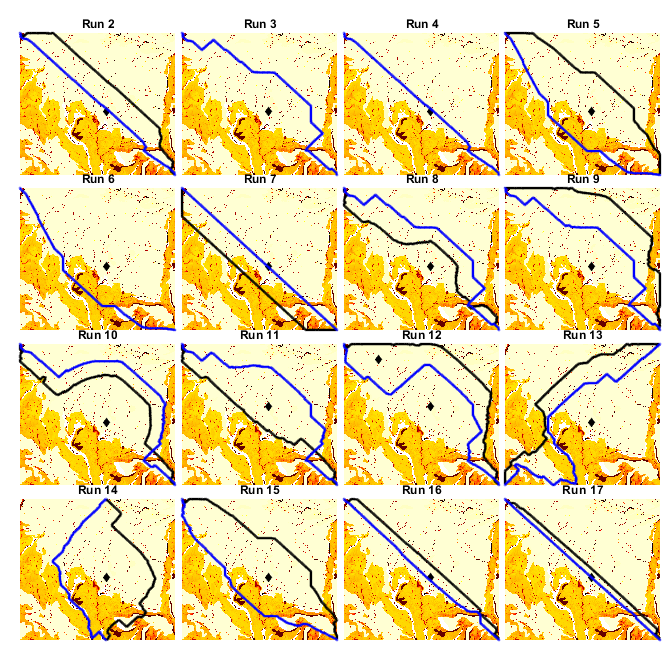}
\caption{Solutions given from subsequent model executions while varying the input parameters from nominal values. The initial route plans are represented by the blue path, while the second planned routes are represented by the black path.}
\label{fig:Results_All}
\end{figure*}

A custom MATLAB script is used to produce the data files representing the optimization model using the software AMPL.  To this end, the MATLAB code first executes the EMT+VS model to determine the soil moisture content of the planning area, then the STRESS model to produce the soil strength needed as an input to the optimization model.  Once the soil strength is determined for each node within the planning area, the MATLAB script will then write the data files that define the sets of nodes and cost parameters necessary for model execution within AMPL.  The AMPL software supports the use of the commercially available solver, CPLEX, which is the specific solver used to solve each model instance of the integer program presented.  The AMPL data and model files can be found in \cite{github_repo}.

For each solution found, the optimization model was first used to determine the optimal path through the terrain without including any historical data (i.e., simulating the first trip through the area of operations).  Then, the initial solution was used as historical data, and the model was solved again to exercise the penalty associated with traveling in close proximity to a recent convoy.  The parameter values used to execute the optimization model can be found in Table~\ref{tab:Parameters_Table}.  Note that a $VCI_{50} = 54$ psi corresponds to the tracked combat engineer vehicle \cite{FM5}.  In order to analyze the sensitivity of the optimization model to changes in the input parameters, a variety of results have been compiled using the STRESS model results.  Table~\ref{tab:Run2-16} summarizes all the model executions with a summary of the outputs of interest.  Figure~\ref{fig:Results_Run1} shows in full detail the terrain analyzed, along with the color scale used for soil strength values, and a legend that summarizes the symbology used for locations of enemy activity, previously traveled routes, and the current route recommendation.  Figure~\ref{fig:Results_All} shows the remaining solutions that omit these details for clarity. A summary of the routes and their respective model parameters used can be found in Table~\ref{tab:Run2-16}.
\begin{enumerate}
    \item \textbf{Run 1:} This route starts at $(0 \mbox{ km},0 \mbox{ km})$ (the northwest corner of the region) and ends at $(9 \mbox{ km},9 \mbox{ km})$ (the southeast corner of the region). It also has a location of known enemy activity at $(5 \mbox{ km},5 \mbox{ km})$.  This run uses the baseline cost function parameters presented in Table \ref{tab:Parameters_Table}.  Because the penalty associated with being close to the location of enemy activity has an exponential decay function, the main decision in the optimization is to weigh the risk of getting stuck on a longer route versus the risk of a kinetic engagement through proximity to prior enemy activity.  When the model is solved again using the initial route as a historical route input, the new route quickly offsets from the previous route but generally follows the original route except when the risk of getting stuck outweighs the risk of being close to a previously traveled route.  As seen in Table \ref{tab:Run2-16}, the majority of the risk in the second route comes from the necessity of crossing multiple areas of poor soil strength.  Secondary risk comes from traversing areas near previously traveled nodes, especially at the start and end of the route.

\end{enumerate}

\begin{table*}[tb]
\centering
\renewcommand{\arraystretch}{1.2}
\caption{Results from varying model parameter values}
\label{tab:Run2-16}
\begin{tabular}{c|c|c|c|c|c|c|c|c|c|c|c|c}
\hline \hline
\textbf{No.} & \textbf{\begin{tabular}[c]{@{}c@{}}VCI\\ {[}psi{]}\end{tabular}} & ${P}$ & ${H}$ & ${R}$ & \textbf{\begin{tabular}[c]{@{}c@{}}Start \\ Point\end{tabular}} & \textbf{\begin{tabular}[c]{@{}c@{}}End \\ Point\end{tabular}} & \textbf{\begin{tabular}[c]{@{}c@{}}Enemy \\ Activity \\ Location\end{tabular}} & \textbf{\begin{tabular}[c]{@{}c@{}}Soil \\ Strength \\ Risk\end{tabular}} & \textbf{\begin{tabular}[c]{@{}c@{}}Historical \\ Route \\ Risk\end{tabular}} & \textbf{\begin{tabular}[c]{@{}c@{}}Enemy \\ Activity \\ Risk\end{tabular}} & \textbf{\begin{tabular}[c]{@{}c@{}}Objective \\ Function\\ Value\end{tabular}} & \textbf{\begin{tabular}[c]{@{}c@{}}Length \\ of Route \\ {[}km{]}\end{tabular}} \\ \hline
\textbf{1} & 54 & 5 & 20 & 20 & (0,0) & (9,9) & (5,5) & 5079 & 1559 & 0.2 & 6639 & 13.8 \\ \hline
\textbf{2} & 54 & 5 & 20 & 0 & (0,0) & (9,9) & (5,5) & 4714 & 1339 & 0 & 6053 & 11.1 \\ \hline
\textbf{3} & 54 & 5 & 0 & 20 & (0,0) & (9,9) & (5,5) & 4282 & 0 & 4.4 & 4287 & 12.4 \\ \hline
\textbf{4} & 54 & 5 & 0 & 0 & (0,0) & (9,9) & (5,5) & 4256 & 0 & 0 & 4256 & 9.8 \\ \hline
\textbf{5} & 54 & 0 & 20 & 20 & (0,0) & (9,9) & (5,5) & 0 & 699 & 5.5 & 705 & 11.7 \\ \hline
\textbf{6} & 54 & 0 & 0 & 20 & (0,0) & (9,9) & (5,5) & 0 & 0 & 5.4 & 5.4 & 11.7 \\ \hline
\textbf{7} & 54 & 0 & 20 & 0 & (0,0) & (9,9) & (5,5) & 0 & 849 & 0 & 849 & 10.8 \\ \hline
\textbf{8} & 54 & 500 & 20 & 20 & (0,0) & (9,9) & (5,5) & 416631 & 3306 & 58.5 & 419996 & 13 \\ \hline
\textbf{9} & 54 & 5 & 2000 & 20 & (0,0) & (9,9) & (5,5) & 13423 & 104450 & 0.1 & 117873 & 14.6 \\ \hline
\textbf{10} & 54 & 5 & 20 & 2000 & (0,0) & (9,9) & (5,5) & 4725 & 2190 & 169 & 7085 & 14.3 \\ \hline
\textbf{11} & 54 & 5 & 20 & 20 & (0,0) & (9,9) & (5,4) & 4695 & 2057 & 50 & 6802 & 11.2 \\ \hline
\textbf{12} & 54 & 5 & 20 & 20 & (0,0) & (9,9) & \begin{tabular}[c]{@{}c@{}}(5,4)\\      (2,1)\end{tabular} & 4749 & 1494 & 112 & 6356 & 14.6 \\ \hline
\textbf{13} & 54 & 5 & 20 & 20 & (9,0) & (0,9) & (5,5) & 2690 & 2026 & 0.6 & 4717 & 13.3 \\ \hline
\textbf{14} & 54 & 5 & 20 & 20 & (5,0) & (5,9) & (5,5) & 4353 & 524.9 & 1.6 & 4879 & 10.2 \\ \hline
\textbf{15} & 35 & 5 & 20 & 20 & (0,0) & (9,9) & (5,5) & 853 & 845 & 4.3 & 1702 & 11.9 \\ \hline
\textbf{16} & 72 & 5 & 20 & 20 & (0,0) & (9,9) & (5,5) & 37686 & 2303 & 238 & 40228 & 9.9 \\ \hline
\textbf{17} & 97 & 5 & 20 & 20 & (0,0) & (9,9) & (5,5) & 156633 & 3056 & 399 & 160088 & 9.6 \\ \hline \hline
\end{tabular}
\end{table*}

\noindent The next six model runs demonstrate the effect of each penalty type by deactivating terms in the objective function (i.e., setting the associated coefficients to zero).

\begin{enumerate}

\setcounter{enumi}{1}

\item \textbf{Run 2:} This run starts at $(0\mbox{ km},0 \mbox{ km})$, ends at $(9 \mbox{ km},9 \mbox{ km})$, and has known enemy activity at $(5 \mbox{ km},5 \mbox{ km})$. However, for this case, the penalty coefficient for movement into nodes close to enemy activity was reduced to zero, which eliminates the contribution from this term in the objective function.  As seen in Figure~\ref{fig:Results_All}, the first solution for this run travels very close to the site of enemy activity. However, the second solution still contains a penalty associated with proximity to a previously traveled route, which causes the second route to parallel the first route except in areas where the soil strength penalty dominates the solution.  

\item \textbf{Run 3:} This run starts at $(0 \mbox{ km},0 \mbox{ km})$, ends at $(9 \mbox{ km},9 \mbox{ km})$, and has a known enemy activity at $(5 \mbox{ km},5 \mbox{ km})$.  Now the coefficient associated with traveling near a previously traveled route is zero.  In this case, it is expected that the first route identified would be identical to the first route in Run 1, and the second route would be the same as the first route because no penalty is associated with prior routes.  This is exactly what is observed in Figure~\ref{fig:Results_All}.

\item \textbf{Run 4:} This run starts at $(0 \mbox{ km},0 \mbox{ km})$, ends at $(9 \mbox{ km},9 \mbox{ km})$, and has known enemy activity at $(5 \mbox{ km},5 \mbox{ km})$.  In this case, the only penalty is associated with the risk of immobilization, as the coefficients for the other two terms are reduced to zero.  Therefore, it is expected that both the first and second routes mimic the first route in Run 2 because this was the only active term for that model solution as well.  This is the behavior that is observed, as seen in Figure~\ref{fig:Results_All}.

\item \textbf{Run 5:} This run starts at $(0 \mbox{ km},0 \mbox{ km})$, ends at $(9 \mbox{ km},9 \mbox{ km})$, and has known enemy activity at $(5 \mbox{ km},5 \mbox{ km})$.  For this run, the coefficient associated with the risk of immobilization from poor soil strength is removed, so it is expected that the solutions will not prioritize traveling in areas of sufficient soil strength.  Instead, they will occupy the fewest number of nodes while also avoiding areas near enemy activity and previously traveled nodes.  The first solution identifies a route that passes through an area of poor soil strength to the south of the enemy activity, and the second route avoids the enemy activity and the previously traveled route by traveling further to the north.

\item \textbf{Run 6:} This run starts at $(0 \mbox{ km},0 \mbox{ km})$, ends at $(9 \mbox{ km},9 \mbox{ km})$, and has enemy activity at node $(5 \mbox{ km},5 \mbox{ km})$.  For this run, only the penalty associated with the proximity to recent enemy activity is considered, as the coefficients on the other two terms are reduced to zero.  Therefore, it is expected that both routes will match the first route from Run 5.  This is exactly the behavior seen in Figure~\ref{fig:Results_All}.

\item \textbf{Run 7:} This run starts at $(0 \mbox{ km},0 \mbox{ km})$, ends at $(9 \mbox{ km},9 \mbox{ km})$, and has known enemy activity at $(5 \mbox{ km},5 \mbox{ km})$. For this case, only the penalty associated with proximity to prior routes is considered, as the coefficients on the other two terms are reduced to zero.  It is expected that the first route will take the shortest path from the starting node to the ending node, and the second route will quickly separate from the first route, then parallel the first route before merging at the end point.  This is the behavior seen in Figure~\ref{fig:Results_All}.

\end{enumerate}

\noindent The next three runs demonstrate the effects of the coefficient magnitudes on the optimization results.

\begin{enumerate}

\setcounter{enumi}{7}

\item \textbf{Run 8:} This run starts at $(0 \mbox{ km},0 \mbox{ km})$, ends at $(9 \mbox{ km},9 \mbox{ km})$, and has known enemy activity at $(5 \mbox{ km},5 \mbox{ km})$.  For this run, the penalty coefficient associated with poor soil strength, $P$, is increased by a factor of 100.  The first route from this run remained unchanged compared to the first route of Run 1, as can be seen in Figure~\ref{fig:Results_Run1} and \ref{fig:Results_All}.  This behavior is expected because the soil strength term already dominates the determination of the first route in Run 1.  However, the second route still shows that the soil strength term dominates, as the crossing point through the area of poor soil strength is nearly the same as the first run because the soil strength consideration now outweighs the risk associated with traveling near the previous route.

\item \textbf{Run 9:} This run starts at $(0 \mbox{ km},0 \mbox{ km})$, ends at $(9 \mbox{ km},9 \mbox{ km})$, and has known enemy activity at $(5 \mbox{ km},5 \mbox{ km})$.  For this run, the penalty coefficient associated with traveling near previous routes, $H$, is increased by a factor of 100.  As can be seen in Figure~\ref{fig:Results_All}, the first route is virtually identical to the first route of Run 8, which is expected because the models are identical when no historical data is considered. However, the second route is pushed a considerable distance from the first route as nodes close to the first route carry heavy penalties.  Because the penalty still decays exponentially with distance from the route, this term does not totally dominate the solution but carries considerable weight.

\item \textbf{Run 10:} This run starts at $(0 \mbox{ km},0 \mbox{ km})$, ends at $(9 \mbox{ km},9 \mbox{ km})$, and has known enemy activity at $(5 \mbox{ km},5 \mbox{ km})$.  For this case, the penalty coefficient associated with proximity to known enemy activity, $R$, is increased by a factor of 100.  As can be seen in Figure~\ref{fig:Results_All}, the first solution travels a clear circular boundary of heavy penalty around the location of enemy activity, which is consistent with the exponential decay of the penalty with increasing distance from the activity.  The first solution is not completely dominated by the proximity to enemy activity, as it is not pushed to the edge of the area of operations.  However, the second solution exhibits a similar route that maintains a large distance from the enemy activity, but it still avoids the prior route due to the associated penalty.

\end{enumerate}

\noindent The next four model runs examine the optimization results when the starting, ending, and enemy activity locations are changed.

\begin{enumerate}

\setcounter{enumi}{10}

\item \textbf{Run~11:} This run starts at $(0 \mbox{ km},0 \mbox{ km})$, ends at $(9 \mbox{ km},9 \mbox{ km})$, and has known enemy activity at $(5 \mbox{ km},4 \mbox{ km})$.  Because the enemy activity was shifted north by 1 km, the optimal routes changed by either moving to the north to maintain an offset from the enemy activity or moving to the south where favorable terrain exists (Figure~\ref{fig:Results_All}). 

\item \textbf{Run~12:} This run starts at $(0 \mbox{ km},0 \mbox{ km})$, ends at $(9 \mbox{ km},9 \mbox{ km})$, and has two locations of enemy activity at $(2 \mbox{ km},1 \mbox{ km})$ and $(5 \mbox{ km},4 \mbox{ km})$.  The first solution maintains a moderate distance from both locations of enemy activity while avoiding the southern region due to poor strength (Figure~\ref{fig:Results_All}).  The second route shifts further northeast to maintain distance from both locations of enemy activity and the previously traveled route.

\item \textbf{Run~13:} This run starts at $(9 \mbox{ km},0 \mbox{ km})$ (northeast corner), ends at $(0 \mbox{ km},9 \mbox{ km})$ (southwest corner), and has known enemy activity at $(5 \mbox{ km},5 \mbox{ km})$.  The first solution exploits a narrow path of high-strength soil that passes through the low-strength area in the southwestern corner of the region.  This route helps mitigate the risk of immobilization from passage through weak soils.  The second route balances the risk associated with the previously traveled route with the risk of getting stuck.  It can be seen from Figure~\ref{fig:Results_All} that the second route is longer and passes through a different narrow region of high soil strength to maintain distance from the first route.

\item \textbf{Run~14:} This run starts at $(5 \mbox{ km},0 \mbox{ km})$ (midpoint of the northern edge), ends at $(5 \mbox{ km},9 \mbox{ km})$ (midpoint of the southern edge), and has known enemy activity at $(5 \mbox{ km},5 \mbox{ km})$. Both routes follow corridors of favorable terrain through the area of poor soil strength to minimize the risk of immobilization.  They also maintain distance from the enemy activity and each other (Figure~\ref{fig:Results_All}).

\end{enumerate}

\noindent The final three runs examine how the solution changes based on the vehicle type.  In all other model runs, the $VCI_{50}$ was 54 psi, which is the value associated with the tracked combat engineer vehicle \cite{FM5}.

\begin{enumerate}

\setcounter{enumi}{14}

\item \textbf{Run 15:} This run starts at $(0 \mbox{ km},0 \mbox{ km})$, ends at $(9 \mbox{ km},9 \mbox{ km})$, and has known enemy activity at $(5 \mbox{ km},5 \mbox{ km})$.  For this model instance, a $VCI_{50}$ value of 35 psi was used, which corresponds to the lighter M2A1 Infantry Fighting Vehicle \cite{FM5}.  In this case, soil strength produces little limitation on the route as the vast majority of nodes have sufficient strength.  Figure~\ref{fig:Results_All} shows that both routes avoid the enemy activity, while the second route also avoids the prior vehicle path.

\item \textbf{Run 16:} This run starts at $(0 \mbox{ km},0 \mbox{ km})$, ends at $(9 \mbox{ km},9 \mbox{ km})$, and has known enemy activity at $(5 \mbox{ km},5 \mbox{ km})$.  For this case, a $VCI_{50}$ value of 72 psi was used, which corresponds to the M51 5 Ton Truck \cite{FM5}.  This value is generally above the weight threshold that can be supported by the soil, so the soil strength term dominates the solution (Figure~\ref{fig:Results_All}).  The optimal routes avoid the enemy activity to some degree and avoid the worst soil conditions, but they generally favor the shortest distance to mitigate the ongoing risk of getting stuck.

\item \textbf{Run 17:} This run starts at $(0 \mbox{ km},0 \mbox{ km})$, ends at $(9 \mbox{ km},9 \mbox{ km})$, and has known enemy activity at $(5 \mbox{ km},5 \mbox{ km})$.  For this run, a $VCI_{50}$ value of 97 psi was used, which corresponds to the M520 8 Ton Cargo Truck \cite{FM5}.  The solutions for this case are completely dominated by the risk of getting stuck.  Unlike Run 4 (where the other two penalty contributions are ignored), two unique routes are identified, but very little offset from the enemy activity and previously traveled route occurs.

\end{enumerate}

%% file: input/Section5.tex
%\xsection{Section 5}\label{sec:section 5}

%Hello

%% file: input/conclusion.tex
\xsection{Conclusion}
\label{sec:conclusion}

This paper presents an optimization-based framework for route planning in open terrain that integrates environmental trafficability constraints with tactical risk considerations into a decision support model. Through the formulation of a risk-based cost function derived from inputs given by the STRESS model, historical convoy movement, and prior enemy activity, this approach extends traditional mobility analysis beyond feasibility and toward route selection based on a multifaceted operational picture. The results of the Maxwell Ranch case study demonstrate how variations in risk weighting influence route geometry and total mission risk, illustrating the flexibility and sensitivity of the model to operational priorities. Overall, the proposed framework provides a structured and quantitative method for evaluating complex trade-offs inherent in open-terrain maneuver planning and establishes a foundation for risk-aware routing applicable to both military and civilian operations.

\subsection{Future Work}
\label{sec:future work}
Future work will extend this model to include slope hazards associated with the terrain.  This will remove arcs within the planning environment, which would enhance the realism of the result while decreasing the model size (which could realize increased performance in the time it takes to discover the optimal solution).  Incorporating another risk factor into the objective function to represent the risk of traveling near civilian populations would also increase the fidelity of the model.  Further analysis on the values selected for the parameters used in the model would also be beneficial to study their effect on the routes provided.  This analysis will allow for the validation of the current values or the opportunity to adjust them to more realistic ones. Also, this work followed a primarily deterministic perspective, so incorporating uncertainty in the risk and other parameters through robust optimization is of interest.

From a computational perspective, conducting mesh and node refinement to identify nodes that cannot be occupied and arcs that cannot be traversed due to logical constraints can further reduce the size of the model, which could improve the tractability as well.  Incorporating a user interface to initialize and run the model would also increase the ease with which the end user can explore the trade-offs and uncertainty in route planning activities. 

Finally, approaching this topic from a holistic point of view through integration of route planning into broader operational capability modeling environments, such as those developed in the Systems Modeling Language (SysML), could further enable the coupling of maneuver planning with other system-level operational analyses.  Together, these efforts will build upon the foundational framework presented in this work and advance the model toward a robust, operationally usable decision-support tool for route planning in complex environments.

%FUTURE WORK:  Steep slopes can be infeasible for travel by military vehicles due to the vehicle's geometry and weight distribution.  Especially in a theater of operations with a high explosive hazards, vehicles tend to be heavy, with a higher center of mass due to their armor, V-shaped hulls, and raised suspensions. Although they are better at deflecting explosions, these vehicles are more susceptible to rollover hazards.  Proximity to civilian populations. Mesh refinement to get rid of impossible arcs.  Because it is impractical to test every section of Earth for soil strength, vehicle routing decisions must account for possible differences between predicted and actual strength.

%\xneed{Future work, UI development for real-time exploration of trade-offs and uncertainty in the route planning activity, integration into operational capability modeling, e.g., with SysML, uncertainty}